\documentclass[%
 aip,
 amsmath,amssymb,
 reprint,%
]{revtex4-1}

\usepackage{graphicx}% Include figure files
\usepackage{dcolumn}% Align table columns on decimal point
\usepackage{bm}% bold math
\usepackage[utf8]{inputenc}
\usepackage[T1]{fontenc}
\usepackage{mathptmx}
\usepackage{etoolbox}
\makeatletter
\def\@email#1#2{%
 \endgroup
 \patchcmd{\titleblock@produce}
  {\frontmatter@RRAPformat}
  {\frontmatter@RRAPformat{\produce@RRAP{*#1\href{mailto:#2}{#2}}}\frontmatter@RRAPformat}
  {}{}
}%
\makeatother
\begin{document}

\preprint{AIP/123-QED}

\title{Implementation of quantum synchronization over a 20-km fiber distance based on frequency-correlated photon pairs and HOM interference}
% Force line breaks with \\
\author{Yuting Liu}
 \affiliation{Key Laboratory of Time and Frequency Primary Standards, National Time Service Center, Chinese Academy of Sciences, Xi’an 710600, China}%Lines break automatically or can be forced with \\
\affiliation{%
School of Astronomy and Space Science, University of Chinese Academy of Sciences, Beijing, 100049, China}
\author{Runai Quan}
 \affiliation{Key Laboratory of Time and Frequency Primary Standards, National Time Service Center, Chinese Academy of Sciences, Xi’an 710600, China}
\affiliation{%
School of Astronomy and Space Science, University of Chinese Academy of Sciences, Beijing, 100049, China}
\author{Xiao Xiang}
 \affiliation{Key Laboratory of Time and Frequency Primary Standards, National Time Service Center, Chinese Academy of Sciences, Xi’an 710600, China}
\affiliation{%
School of Astronomy and Space Science, University of Chinese Academy of Sciences, Beijing, 100049, China}
\author{Huibo Hong}
 \affiliation{Key Laboratory of Time and Frequency Primary Standards, National Time Service Center, Chinese Academy of Sciences, Xi’an 710600, China}
\affiliation{%
School of Astronomy and Space Science, University of Chinese Academy of Sciences, Beijing, 100049, China}
\author{Tao Liu}
\affiliation{Key Laboratory of Time and Frequency Primary Standards, National Time Service Center, Chinese Academy of Sciences, Xi’an 710600, China}
 \affiliation{School of Astronomy and Space Science, University of Chinese Academy of Sciences, Beijing, 100049, China}%Lines break automatically or can be forced with \\
\author{Ruifang Dong}%
\altaffiliation[Electronic mail: ]{dongruifang@ntsc.ac.cn}
%\email{dongruifang@ntsc.ac.cn}
\affiliation{Key Laboratory of Time and Frequency Primary Standards, National Time Service Center, Chinese Academy of Sciences, Xi’an 710600, China}
\affiliation{School of Astronomy and Space Science, University of Chinese Academy of Sciences, Beijing, 100049, China}
\author{Shougang Zhang}
 %\homepage{http://www.Second.institution.edu/~Charlie.Author.}
\affiliation{Key Laboratory of Time and Frequency Primary Standards, National Time Service Center, Chinese Academy of Sciences, Xi’an 710600, China}
\affiliation{%
School of Astronomy and Space Science, University of Chinese Academy of Sciences, Beijing, 100049, China}
\date{\today}% It is always \today, today,
             %  but any date may be explicitly specified

\begin{abstract}
The quantum synchronization based on frequency-correlated photon pairs and HOM interference has shown femtosecond-level precision and great application prospect in numerous fields depending on high-precision time-frequency signals. Due to the difficulty of achieving stable HOM interference fringe after long-distance fiber transmission, this quantum synchronization is hampered from long-haul field application. Utilizing segmented fibers instead of a single long-length fiber, we successfully achieved the stable observation of the two-photon interference of the lab-developed broadband frequency-correlated photon pairs after 20 km-long fiber transmission, without employing auxiliary phase stabilization method. Referenced to this interference fringe, the balance of the two fiber arms is successfully achieved with a long-term stability of 20 fs. The HOM-interference-based synchronization over a 20-km fiber link is thus demonstrated and a minimum stability of 74 fs has been reached at 48,000 s. This result not only provides a simple way to stabilize the fiber-optic two-photon interferometer for long-distance quantum communication systems, but also makes a great stride forward in extending the quantum-interference-based synchronization scheme to the long-haul field applications.
\end{abstract}

\maketitle
\section{Introduction}
The two-photon Hong-Ou-Mandel (HOM) interference\citep{HOM1987} has been widely utilized in many applications of quantum technologies, such as high-precision measurement\citep{Attosecond2018,Simultaneous2000,Tests1999}, quantum teleportation\citep{Highly2015,QuantumTeleportation2021}, and so on. Based on the HOM interference, quantum clock synchronization (QCS) has been also proposed and experimentally demonstrated as an advantageous method to the classical synchronization schemes. Since the first report of the QCS experiment based on frequency-correlated-photon pairs and HOM interference on a 4 km fiber link, which demonstrated a minimum timing stability of 0.44 ps at the averaging time of 16,000 s\citep{Demonstration2016}, we further improved the experiment by realizing a minimum timing stability of 60 fs at 25,600 s on a 6 km fiber link\citep{Simulationrealization2019}. Furthermore, we provided a quantitative characterization of the achievable timing accuracy and stability, which has been verified by the experimental results.  \par
Analogous to other quantum communication systems, the key of realizing this fiber-based long-distance QCS is to stably achieve the HOM interference fringe after fiber transmission. Because of the surrounding environment variation (such as temperature, humidity, mechanical and acoustic disturbances and so on), the effective path difference between the two arms after long-distance fiber transmission will accumulate significant drift that surpasses the two-photon coherence length. Therefore, efforts to reduce the fiber path drift within the coherence length of the photon pairs become a challenge. The typical methods include applying narrowband spectral filter to increase the two-photon coherence time\citep{Photonbunching2005} or employing continuous optical length control based on an auxiliary reference light\citep{Active2016}. However, the narrowband filter will not only induce loss but also degrade the spectral correlation between the photon pairs, which is key to various quantum information applications. While the latter method will add both complexity and background noise into the setup as a slight portion of the reference light might enter the single-photon detectors as well.\par
In this letter, we present a new method of effectively reducing the fiber path drift by utilizing segmented fibers instead of a single long-length fiber. In this way, we successfully achieve the stable observation of HOM interference after 20 km fiber transmission on both arms. By using the lab-developed telecom-wavelength broadband frequency-correlated photon pairs as the source\citep{Characterization2015}, the HOM interference after the fiber transmission were measured with a visibility of 60 $\%$ and width of 3.25 ps, which well verified the dispersion-immune property of HOM interference\citep{Photonbunching2005,Characterization2015,Entangled2002}. Based on this HOM interference fringe, the two fiber arms were then stably balanced via a feedback loop applied onto a variable delay in one arm, with a long-term time deviation (TDEV) of about 20 fs. With this setup, the quantum synchronization over a 20-km fiber distance is thus demonstrated, achieving a minimum stability of 74 fs at 48,000 s. The absolute synchronization accuracy is measured to be 50.5 $\pm$ 7.2 ps, which agrees well with the theoretical simulation of 44.14 ps. These results provide us a bright prospect of applying the HOM-interference-based synchronization scheme into the long-haul field fiber links.

\section{Theoretical mechanism}
According to the QCS protocol based on the HOM interference\citep{ClockSynchronization2004}, it is assumed that clocks A and B are sufficiently stable and have the same rate with respect to the coordinate time. Once the paths between the two clock sites to baseline are balanced, the coordinate time of photon arriving at sites A and B are equal. Consequently, the time offset acquired by the second-order coincidence measurement is the time offset between clocks A and B. In the protocol, the path balance can be achieved by constructing a HOM interferometer, and the balance point can be readily found by the minimum coincidence of HOM interference.

To analyze the effect of the induced fiber path drift in the HOM interference, the theoretical model in Ref.[8] is used. Assume the fiber lengths of the two arms are both equal to $l$, the remaining path-delay difference can be described by:
\begin{small}
\begin{equation} \label{eq:t}
\tau=\left(k_{i, f}^{\prime}-k_{s, f}^{\prime}\right) l-\left(k_{i, f}^{\prime \prime} \omega_{0, i}-k_{s, f}^{\prime \prime} \omega_{0, s}\right) l
\end{equation}
\end{small}

Here, $k_{s(i),f}$ denotes the wavenumbers for the signal ($s$) and idler ($i$) photons in the fiber channel, $k_{s(i), f}^{\prime}$ and $k_{s(i), f}^{\prime\prime}$ represent the first- and second-order derivatives of the wavenumbers around the center frequency $\omega_{0, s(i)}$. Due to the sensitivity of the refractive index and thermal expansion of the optical fiber to various environmental parameters such as temperature fluctuation, the path-delay difference given by the above Eqn.(\ref{eq:t}) is changed and can be expressed as
\begin{small}
\begin{equation}
\Delta \tau=\sqrt{\frac{\partial^2\left(k_{i, f}^{\prime} l\right)}{\partial T^2}+\frac{\partial^2\left(k_{s, f}^{\prime} l\right)}{\partial T^2}+\frac{\partial^2\left(k_{i, f}^{\prime \prime} \omega_{0, i} l\right)}{\partial T^2}+\frac{\partial^2\left(k_{s, f}^{\prime \prime} \omega_{0, s} l\right)}{\partial T^2}}\Delta T 
\end{equation}
\end{small}

\begin{figure}
\includegraphics[width=0.45\textwidth]{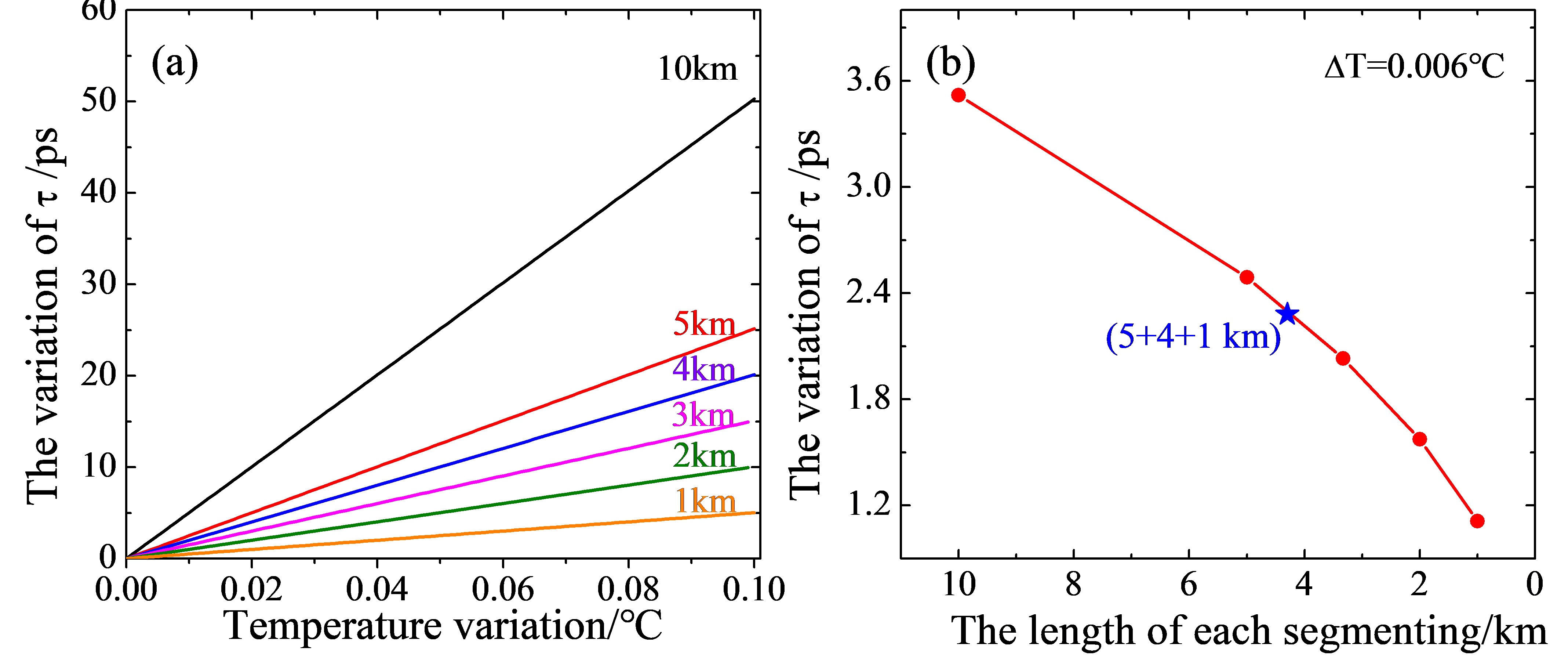}% Here is how to import EPS art
\caption{\label{fig:Simulation} Simulation of the (a) path-delay variation as a function of the temperature when the single fiber length is chosen as 1, 2, 3, 4, 5, and 10 km, respectively and (b)  the influence of dividing the 10 km-long single fiber into segments by the path-delay variation as a function of the length of each segment. In the simulation, the  spectral parameters of the lab-developed frequency-correlated photon pair source and the refractive index equation of the optical fiber are used. The ambient temperature fluctuation is reasonably chosen as 0.006 $^{\circ}$C, and the blue star in the figure represent the fiber combination used in the experiment.}
\end{figure}

Neglecting the thermal expansion of the fiber length, it can be simplified to 
\begin{equation}
\Delta \tau=B l\Delta T
\end{equation}
where 
\begin{small}
\begin{equation}
B=\sqrt{\frac{\partial^2\left(k_{i, f}^{\prime}\right)}{\partial T^2}+\frac{\partial^2\left(k_{s, f}^{\prime}\right)}{\partial T^2}+\frac{\partial^2\left(k_{i, f}^{\prime \prime} \omega_{0, i}\right)}{\partial T^2}+\frac{\partial^2\left(k_{s, f}^{\prime \prime} \omega_{0, s}\right)}{\partial T^2}}
\end{equation}
\end{small}represents the contribution factor from temperature sensitivity of the refractive index of the fiber. In our experiment, the mean center wavelengths of the signal and idler photons are measured to be 1574.4 nm and 1574.7 nm, respectively. Combined with the refractive index equation of the single mode fiber\citep{Hedekvist201217TA}, $B$ is deduced as 5.03 $\times$ $10^{-14}$ s/(m$\cdot$$^{\circ}$C) according to the above theory when the operation temperature is set to 22$^{\circ}$C . The variation of $\tau$ with respect to a temperature fluctuation reaching 0.1 $^{\circ}$C is shown in FIG. \ref{fig:Simulation}(a). It can be seen that the variation has a proportional dependence on the single fiber length (1, 2, 3, 4, 5, and 10 km).

From Eqn.(3) it can also be inferred that, by dividing $l$ into several segments of $l_{1}, \ldots, l_{m}$, where $l=\sum_{k=1}^{m} l_{k}$, the path-delay difference variation can be reduced to
\begin{equation}
\Delta \tau^{\prime}=B \sqrt{\sum_{k=1}^{m} l_{m}^{2}}\Delta T
\end{equation}

To show the influence of fiber segmenting on the path-delay difference variation, we consider the cases that a 10-km long optical fiber are evenly divided into 1, 2, 3, 5, and 10 segments, respectively. As the temperature of our laboratory is controlled at 22 $^{\circ}$C with a periodic fluctuation of $\pm$ 0.5 $^{\circ}$C within one hour, a temperature variation of $\sim$ 0.006 $^{\circ}$C during a HOM measurement period lasting for 20 seconds is deduced. The relationship between the path-delay difference variation and the length of individual segmented fiber is shown in FIG. \ref{fig:Simulation}(b). It can be readily seen that the variation induced by a single 10 km fiber has exceeded the two-photon coherence time of the utilized frequency-correlated photon pair source (measured as $\sim$3.25 ps in our experiment), and this well explains why a HOM dip could not be observed with the single 10-km fiber in the current configuration. On the other hand, dividing the 10 km-long single fiber into segments can reduce effectively the path-delay variation. Note should be taken that, segmenting the transfer fiber may lead to extra losses as more fiber connectors should be used and there will be stress at the joint of each segment of optical fiber that influence the HOM interference. Considering the tradeoff between the path delay variation and extra loss, in the following demonstration of HOM-interference-based quantum synchronization, three segments of optical fiber combination (5 km + 4 km + 1 km) are utilized. According to the simulation, the variation can be reduced to a value of 2.28 ps (marked with blue star in FIG. \ref{fig:Simulation}(b)), which is well below the two-photon coherence time. 

\section{Experimental setup}
\begin{figure*}
\includegraphics[width=0.7\textwidth]{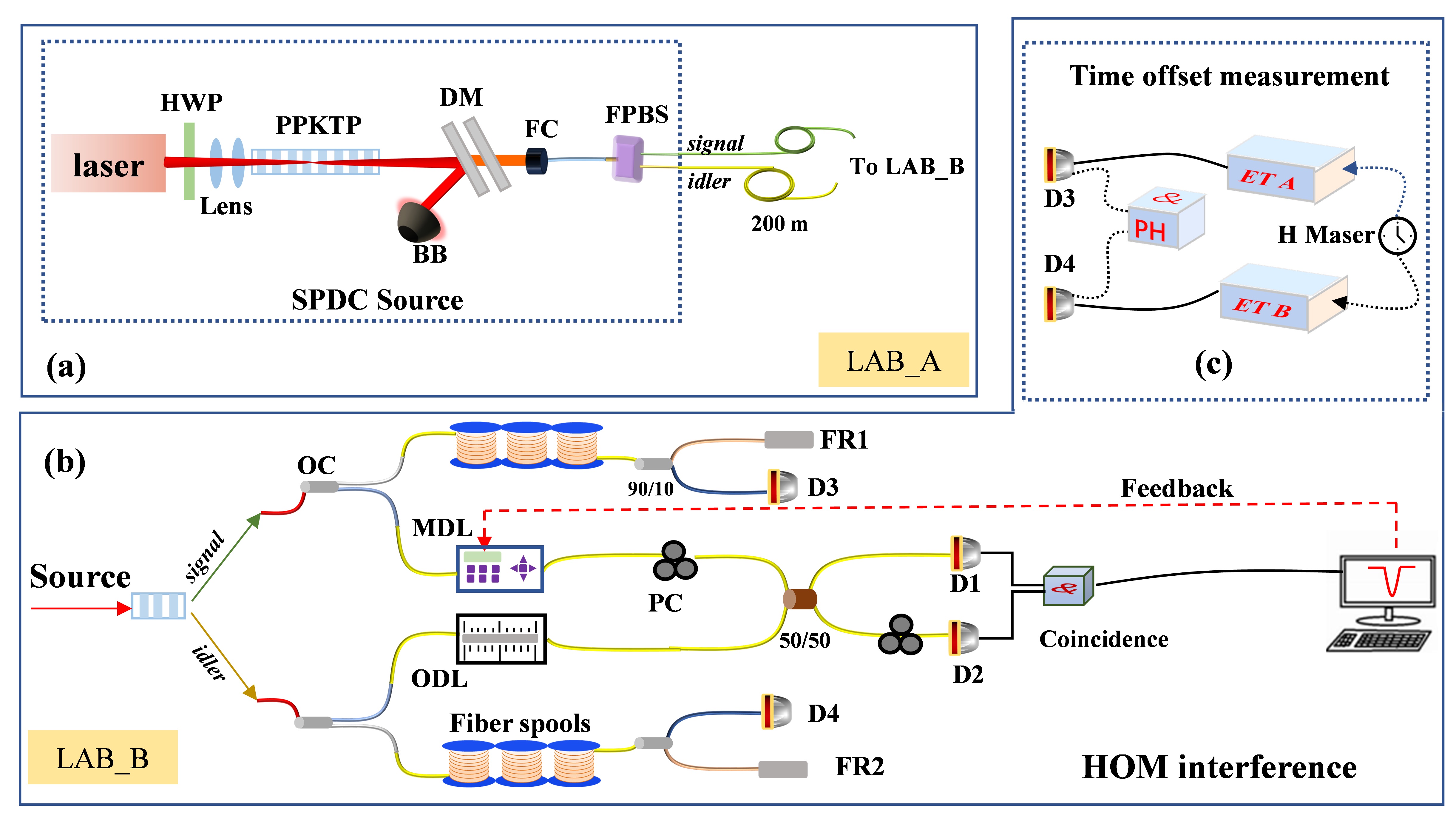}% Here is how to import EPS art
\caption{\label{fig:setup} Schematic of the experimental setup. (a) the generation process of frequency correlated entangled source. HWP: half wave plates; DM: dichroic mirrors; BB: beam block; FC: fiber coupler; FPBS: fiber polarization beam splitter; (b) the path-balance feedback loop based on HOM interferometer. MDL: motorized optical delay line; OC: optical circulators; ODL: manual optical delay line; fiber spools: 5 km + 4 km + 1 km, 90/10 $\&$ 50/50: beam splitters; D1, D2, D3 and D4 are SNSPDs; PC: polarization coupler; FR: Faraday rotators; (c) the time offset measurement. PH: PicoHarp 300; ETA and ETB: Even timers A and B.}
\end{figure*}

The experimental setup of the HOM-interference-based QCS over a 20 km optical fiber link is shown in FIG. \ref{fig:setup}. A broadband frequency-correlated photon pair source\citep{Simulationrealization2019,Characterization2015}(FIG. \ref{fig:setup}(a)) was generated via the process of spontaneous parametric down conversion (SPDC), in which a 20 mm PPKTP  (type II) with a poling period of 46.146 $\mu$m was pumped by a 787 nm femtosecond laser with the spectral bandwidth of 25 nm and the repetition rate of 75 MHz (Fusion 20-150, FemtoLasers). Through a half wave plate (HWP) and a fiber polarization beam splitter (FPBS), the entangled signal and idler photons located at $LAB$\_$A$ were separated and sent to two 200 m-long outdoor fibers, and arrived at another laboratory ($LAB$\_$B$). 

Then they were transmitted through two 10 km-long commercial single-mode optical fibers, each including three segments of fiber spools (5 km + 4 km + 1 km) (Yangtze Optical Fiber and Cable Company Ltd.). At the fiber transmission end, each photon beam was divided into two parts by a 90/10 single-mode fiber beam splitter (FBS): 90 $\%$ of the photons were reflected by Faraday rotators and backtracked into the HOM interferometer for achieving path-balance locking (as shown in FIG. \ref{fig:setup}(b)); the remaining 10 $\%$ of the photons were detected directly by the superconductive nanowire single photon detectors (SNSPDs, Photon Technology Co., Ltd.) D3 and D4 for the time offset measurement between the two clocks (as shown in FIG. \ref{fig:setup}(c)). In the path-balance part, the adjustable optical delay lines were composed of a motorized optical delay line (MDL) with a resolution of 1 fs (MDL-002, General Photonics Inc.) for scanning the optical path-length delay, and a manual optical delay line (ODL) (General Photonic Inc.), which were used for compensating the path discrepancy caused by MDL. Two fiber polarization controllers (PC) were used to match the polarization modes of the propagating photons. The signal and idler were directed to a 50/50 fiber-optical beam splitter (BS) and detected by the SNSPDs (D1 and D2). In the time offset measurement part, the output of D3 and D4 were delivered to two even timers (ET, A033-ET, Eventech Ltd.) for arrival time tagging and recording, which were both referenced to a H-maser. Through the second-order correlation calculation of the ET results\citep{Highprecision2020}, the absolute time offset could be obtained. For comparison, a time-correlated single-photon counting system (TCSPC, PicoHarp 300, PicoQuant Inc.) was also used for proof-of-principle QCS, where the clicks of the two detectors are regarded as the virtual clocks.

\section{Results and analysis}
\begin{figure}
\includegraphics[width=0.25\textwidth]{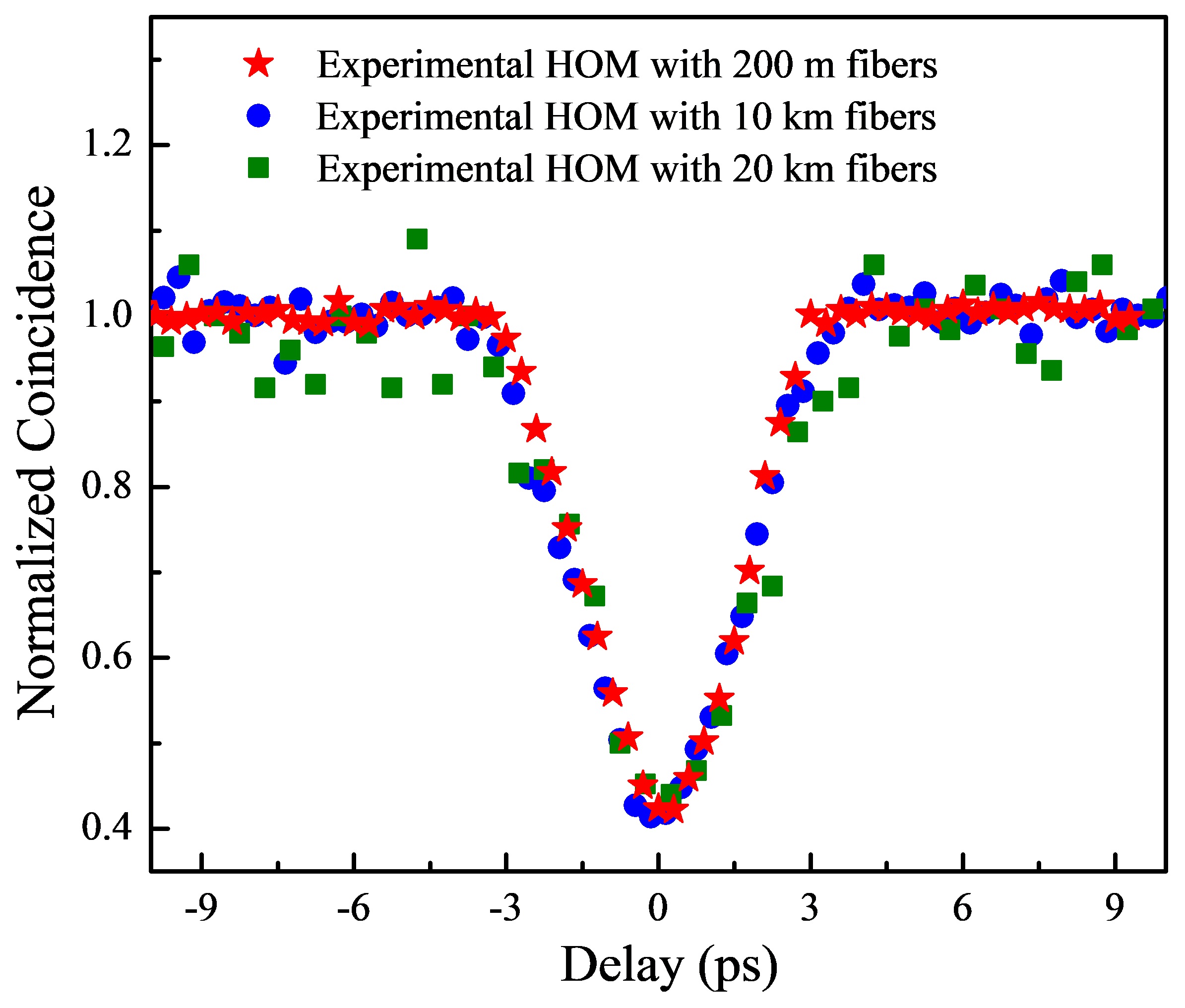}% Here is how to import EPS art
\caption{\label{fig:HOM} Measurement of the HOM interferogram over different optical fiber lengths: red pentagrams, blue circles and green squares represent the experimental results with 200 m,10 km and 20 km, respectively.}
\end{figure}

Measurement of the HOM interferogram over 20 km-long optical fiber (combination of two 5 km + 4 km + 1 km fiber spools) is the first stage of the QCS experiment, which is shown by green squares in FIG. \ref{fig:HOM}. The observed HOM interferometric dip yields a visibility of 60 $\%$ and a coherence time of 3.25 ps. When the faraday rotators (FR) on each fiber path and the long-distance fiber were removed, the measured HOM interferograms of 10 km and 200 m fiber links are also depicted in FIG. \ref{fig:HOM} in blue circles and red pentagrams, respectively. It is clearly shown that the visibilities and coherence widths are almost the same in all three cases, which is consistent with the conclusion in Ref [12, 16], that the width of the interference dip is insensitive to the second-order dispersion contribution.

\begin{figure}
\includegraphics[width=0.48\textwidth]{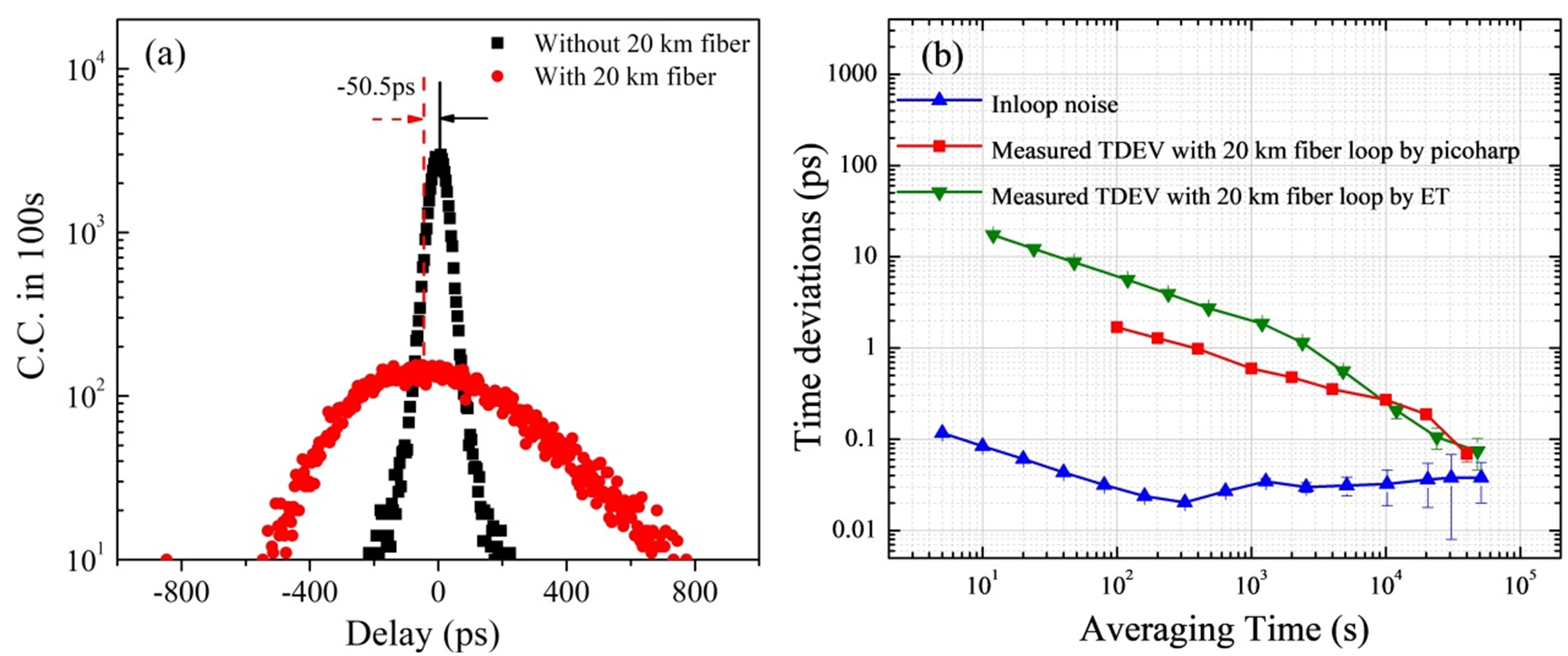}% Here is how to import EPS art
\caption{\label{fig:clock synchronization} (a) Measured time correlation distributions between two single-photon detectors at an acquisition time of 100 s with (in red circles) and without (in black squares) the two, 10 km fiber links in the setup. (b) TDEV results for the in-loop timing jitter with the two 10 km-long fiber links balanced (blue upward triangles), the measured time offsets over the 20 km fiber distance acquired by PicoHarp 300 (red squares) and A033-ET (green downward triangles).}
\end{figure}

It is necessary to point out that the HOM interferometer setup based on the single 10-km fiber spools was initially used, but no HOM interferogram was observed. The reason has been theoretically analyzed  in Section II, and it could be attributed to temperature-dependent path difference drift between the two arms exceeds the coherence length determined by the two-photon spectral bandwidth. By utilizing segmented fibers instead of a single long-length fiber to construct the 20 km-long fiber link, the HOM interferogram is clearly observed. Based on this HOM interference fringe, the two fiber arms were then stably balanced via a one-step feedback loop applied onto the MDL. As shown by the blue upward triangles in FIG. \ref{fig:clock synchronization}(b), and a long-term time deviation (TDEV) of about 20 fs.

With the path-balancing system enabled with the HOM dip stabilization, the time offset between clocks A and B can be acquired by the second-order coincidence measurement. Taking the advantage of unique nonlocal dispersion cancellation effect inherent in frequency-correlated photon pair source, the coincidence of the photon pairs after passing through similar dispersion elements will experience a minimum  temporal width broadening\citep{Quantification2020}, therefore there is no need for extra dispersion compensation to obtain higher synchronizing stability. 

FIG. \ref{fig:clock synchronization}(a) shows the measured time correlation distributions at an acquisition time of 100 s with (in red circles) and without (in black squares) the two, 10 km-long segmented fiber spools using PicoHarp. By Gaussian fitting of the coincidence distributions, the time offset with respect to the maximum coincidences as well as the width can be extracted. Compared with the measured time offset without the fiber spools, an absolute time offset shift of 50.5 $\pm$ 7.2 ps was obtained that is in accordance with the theoretical simulation of 44.14 ps\citep{Simulationrealization2019}. The coincidence width was broadened from 62 $\pm$ 0.37 ps to 514.8 $\pm$ 3.52 ps through the long-distance fiber, which is consistent with the theoretical simulation when the signal and idler photons transmit over 10 km fiber separetely. The synchronizing stability in term of TDEV as a function of the averaging time within 20 km fiber links is shown as FIG. \ref{fig:clock synchronization}(b) by red squares. A minimum time stability of 1.8 ps was achieved at an averaging time of 100 s.

Subsequently, to build a scenario close to field application, we utilized two Event timers (ETs) triggered by 10 MHz and 1 PPS clock signal from the same H-maser to measure the time offset of 20 km fiber links. The input photon rate for each ET was 12 kcps due to its intrinsic acquisition rate limit, the acquisition period was set as 12 s. The TDEV measured by ETs is displayed in FIG. \ref{fig:clock synchronization}(b) in green downward triangles. It's clear to see that the time stability measured by PicoHarp was better than that measured by ETs in the short term. Such discrepancy comes from the coincidence count rates in the two cases, which are were 272.54 $\pm$ 1.96 cps by PicoHarp and 35.46 $\pm$ 1.89 cps by ETs, respectively. From the perspective of long-term timing stability, both TDEV curves eventually approach the time jitter of the system itself (in blue upward triangles), which are independent of the transmission path and measurement device\citep{Smotlacha2010TimeTI}. At an averaging time of 48,000 s, a minimum TDEV of 74 fs was reached. These series of experimental results provide an important basis for the realization of clock synchronization between inter-metropolitan areas based on second-order quantum coherence of entangled photons.

\section{conclusion}
In this letter, we demonstrate a new strategy of employing fiber segments instead of a single fiber to stably acquire the HOM interferogram of a broadband frequency correlated photon pairs after 20 km-long fiber transmission. The HOM interference visibility was measured as 60 $\%$ with a dip width of 3.25 ps,  which well verified the dispersion-immune property of HOM interference.  On this basis, we successfully achieved the second-order QCS over a 20 km fiber distance and got a minimum synchronizing stability of 74 fs at the 48,000 s with a synchronization accuracy of 50.5 $\pm$ 7.2  ps. The above research breaks through the limitation of long-distance transmission and provides a new idea for the development of a long-distance fiber-based quantum transmission. The results lay a foundation for the realization of QCS towards inter-metropolitan optical network based on this algorithm.

\begin{acknowledgments}
This work was supported by the National Natural Science Foundation of China (Grant Nos. 12033007, 61875205, 61801458, 91836301), the Frontier Science Key Research Project of Chinese Academy of Sciences (Grant No. QYZDB-SW-SLH007), the Strategic Priority Research Program of CAS (Grant No. XDC07020200), the “Western Young Scholar” Project of  CAS (Grant Nos. XAB2019B17 and XAB2019B15),  the Chinese Academy of Sciences Key Project (Grant No. ZDRW-KT-2019-1-0103) and the Youth Innovation Promotion Association, CAS (Grant No. 2021408).
\end{acknowledgments}

\nocite{*}
\bibliography{aipsamp}% Produces the bibliography via BibTeX.

\end{document}